# Origin and spectroscopic determination of trigonal anisotropy in a heteronuclear SMM

L. Sorace,[a,*] M.-E. Boulon,[a] P. Totaro,[a] A. Cornia,[b] J. Fernandes-Soares,[c] R. Sessoli[a]

[a] *Dipartimento di Chimica Ugo Schiff and INSTM RU, Università degli Studi di Firenze, Via della Lastruccia 3-13, 50019 Sesto Fiorentino-FI, Italy. E-mail: lorenzo.sorace@unifi.it; Fax: +0039 055 4573372; Tel: +0039 055 4573336*
[b] *Dipartimento di Scienze Chimiche e Geologiche and INSTM RU, Università degli Studi di Modena e Reggio Emilia, Via G. Campi 183, 41125 Modena, Italy*
[c] *Departamento de Química, Universidade Federal do Paraná, Centro Politécnico, 81530-900 Curitiba-PR, Brazil.*

W-band ($\nu \cong 94$ GHz) Electron Paramagnetic Resonance (EPR) spectroscopy was used for a single-crystal study on a star-shaped $Fe_3Cr$ Single Molecule Magnet (SMM) with crystallographically-imposed trigonal symmetry. The high resolution and sensitivity accessible with W-band EPR allowed us to determine accurately the axial zero-field splitting terms for the ground ($S=6$) and first two excited states ($S = 5$ and $S = 4$). Further, spectra recorded by applying the magnetic field perpendicular to the trigonal axis showed a $\pi/6$ angular modulation. This behavior is a signature of the presence of trigonal transverse magnetic anisotropy terms which were spectroscopically determined for the first time in a SMM. Such an in-plane anisotropy could only be justified by dropping the so-called "giant spin approach" (GSA), and by considering a complete multi-spin approach (MSA). From a detailed analysis of experimental data with the two models, it emerged that the observed trigonal anisotropy directly reflects the structural features of the cluster, *i.e.* the relative orientation of single-ion anisotropy tensors and the angular modulation of single-ion anisotropy components in the hard plane of the cluster. Finally, since high-order transverse anisotropy is pivotal in determining the spin dynamics in the quantum tunneling regime, we have compared the angular dependence of the tunnel splitting predicted by the two models upon application of a transverse field (Berry Phase Interference).

## I. INTRODUCTION

Polynuclear transition metal complexes provided, a couple of decades ago, the first examples of individual magnetic molecules exhibiting a memory effect at low temperature.[1–3] Since then, the family of Single Molecule Magnets (SMMs) has grown considerably and now includes complexes of lanthanides and actinides, as well as a number of mononuclear systems.[4–7] Polynuclear SMMs based on transition metals typically exhibit a large spin ($S$) ground state which stems from intramolecular superexchange interactions between the constituent metal ions and is accompanied by an easy-axis type anisotropy. The two ingredients afford an energy barrier to reorientation of the magnetic moment and result, under favorable conditions, in a memory effect. The identical structure of each molecule in a crystal, together with the shielding provided by the ligand shell that surrounds the magnetic core, make these systems ideal testing grounds for studying quantum phenomena in nanoscale magnets. Indeed, quantum tunneling (QT) of the magnetization and quantum phase interference have been reported for the first time in SMM systems.[8–10] Of paramount importance for the appearance of quantum effects is transverse magnetic anisotropy. In fact, when the anisotropy is purely axial, *i.e.* the plane perpendicular to the easy axis is completely isotropic, QT is forbidden in a longitudinal field.[11] By contrast, transverse anisotropy can mix spin states localized on different sides of the barrier, thereby opening effective tunneling pathways. At fields where two levels would otherwise cross, level repulsion takes place and the resulting tunnel splitting (TS) is directly related to the magnetization tunneling rate through Landau-Zener-Stückelberg formula.[12–14] The low temperature magnetic properties of such systems are usually analyzed using a giant-spin approach (GSA). Within this formalism, only the ground spin multiplet is considered and $S$ is treated as an exact quantum number. Magnetic anisotropy is then introduced as a perturbation acting on the ground manifold and is described using a multipolar series expansion with terms up to $2S$-th order in spin operators, the so-called Stevens operator equivalents, $B_k^q \hat{O}_k^q$.[15] The main advantage of the GSA lies in the relatively small number of free parameters required, since the number of terms is both spin ($k \leq 2S$, $-k \leq q \leq k$) and symmetry restricted. Furthermore, the spin Hamiltonian matrix has dimensions $(2S+1)\times(2S+1)$ only and can be easily diagonalized even in cases where the complete Hilbert space of the multispin system is unmanageably large. To correctly grasp the origin of high-order ($k > 2$) anisotropy terms appearing in the GSA, as well as to account for some subtle effects in relaxation, it is essential to adopt a multi-spin approach (MSA) which explicitly considers the internal degrees of freedom, e.g. the anisotropy of each constituent spin and the details of spin-spin interactions.[16–19] These high-order anisotropies are especially relevant in axially symmetric molecules, where second-order ($k = 2$) transverse anisotropy ($q \neq 0$) vanishes and QT can be promoted only by transverse terms with $k > 2$ and $q \neq 0$.

In striking contrast with their aforementioned importance, high-order transverse anisotropies have been experimentally determined, and their relation to the multispin nature of the systems proved, only for two SMMs with fourfold symmetry.[18,19] Thus, for $Mn_{12}^tBuAc$ (a derivative of the archetypal SMM $Mn_{12}Ac$), some of us have shown that the GSA requires the inclusion of sixth-order terms which can be traced back to the tilting of the single-ion easy axes.[19] A similar approach was applied to a tetranickel(II) cluster, which could be treated exactly due to the small dimension of its Hilbert space (81×81).[18]

An interesting advance in this field would be the analysis of systems with rigorous threefold (trigonal) symmetry. Indeed, since $\hat{O}_k^q$ terms with $q = 3,6$ couple only states differing by $\Delta M_S = \pm 3, \pm 6$ where $M_S$ labels the projection of the total spin onto the $C_3$ ($z$-) axis, nonzero tunneling gaps would be limited to level crossings with $|\Delta M_S| = 3n$ ($n$ positive integer). This should provide a peculiar periodicity of the TS, and thus of magnetization dynamics, on application of a transverse field.[20] However, despite the relevant number of threefold symmetric SMMs so far isolated, no spectroscopic determination of their $B_k^q$ ($q = 3, 6$) parameters is available in the literature.[21–26] In some earlier reports, small departures from threefold symmetry had to be assumed to explain available relaxation data,[27] or high-order transverse anisotropies were only roughly estimated.[28] More recently, Del Barco *et al.* found the signature of threefold symmetry in the low temperature quantum relaxation of a trimanganese(III) SMM.[29] Here, the TS dependence on transverse field was apparently independent on the field orientation, owing to the small magnitude of the trigonal anisotropy.

Among SMMs with potential threefold symmetry, a most notable place is occupied by the star shaped tetrairon(III) ($Fe_4$) derivatives, which have shown a unique combination of structural and electronic robustness and ease of functionalization.[23,30–38] The size of their Hilbert space (1296 × 1296) is small enough to enable a detailed treatment of their electronic structure and spin dynamics using a MSA.[39,40] Furthermore, some $Fe_4$ derivatives such as $[Fe_4(L)_2(dpm)_6]$ and $[Fe_4(L')(EtO)_3(dpm)_6]$ have crystallographically imposed threefold symmetry ($H_3L = Me-C(CH_2OH)_3$; $H_3L' = {}^tBu-C(CH_2OH)_3$; Hdpm = dipivaloylmethane).[23,40] However, preliminary single crystal Electron Paramagnetic Resonance (EPR) studies at W-band ($\nu \cong 94$ GHz) on such clusters failed to reveal a reliable angular dependence of the resonance fields in the hard plane.[40,41]

In an effort to synthesize heterometallic clusters with the same topology, in the past few years we have devised a procedure to replace the central iron(III) with a different tripositive metal ion M. The first synthetic method we described was based on a one-pot reaction and resulted in a solid solution of $Fe_3M$ and $Fe_4$.[42,43] Much better suited for detailed spectroscopic studies are samples prepared through a three-step synthetic approach, which reduces the amount of $Fe_4$ to below the detection limit.[44] Of particular interest among these new heterometallic systems are $Fe_3Cr$ complexes characterized by an $S = 6$ ground state and an axial zero field splitting (zfs) parameter $D \cong -0.18$ cm$^{-1}$. These parameters are ideal for EPR investigations using commercial W-band spectrometers, since the whole spectrum can be observed within 6 T in any orientation. Moreover, in the case of an $S = 6$ state in trigonal symmetry the two states of the ground doublet, *i.e.* $M_S = \pm 6$, are directly admixed by transverse anisotropy, since $\Delta M_S = 12 = 3n$. We have thus synthesized the new complex $[Fe_3Cr(L)_2(dpm)_6]$ (**1**) and found that it has crystallographically-imposed trigonal ($D_3$) symmetry, like its tetrairon(III) analogue.[23,40] We present here a single-crystal W-band EPR study on **1**, which has provided the first spectroscopic determination of high-order transverse anisotropy in a threefold symmetric SMM. The results allow us to draw a detailed picture of the relation between GSA and MSA, highlighting the role of non-collinear single-ion anisotropy tensors. Based on the obtained parameter set, we finally provide a useful prediction concerning the angular dependence of the TS and of the low temperature spin dynamics in these systems.

## II. EXPERIMENTAL SECTION

*Synthesis*. The synthesis of **1** as a pure $Fe_3Cr$ phase followed the procedure reported in Ref. [44] using $H_3L$ = 2-hydroxymethyl-2-methyl-propane-1,3-diol in place of 2-hydroxymethyl-2-ethyl-propane-1,3-diol, with recrystallization by slow evaporation of a *n*-hexane solution. The efficiency of this recently-developed synthetic route for the isolation of pure heterometallic species was here further confirmed by measuring the *ac* susceptibility of a powder sample of **1** in the range 0.03–60 kHz with a home developed probe based on the Oxford Instruments MAGLAB platform.[45] The results revealed a unique peak in the $\chi''$ vs. frequency plots, (See Figure S1 in supplementary material[46]), measured between 1.7 and 5 K, with no detectable contribution from $Fe_4$ species. The effective energy barrier for the reversal of the magnetization extracted from the Arrhenius plot is indeed $U_{eff}/k_B$ = 12.1 ± 0.1 K, hence similar to that of previously reported $Fe_3Cr$ derivatives,[42–44] while the corresponding $Fe_4$ derivative has an energy barrier of 17.0 K.[40]

*X-ray Diffractometry*. Freshly synthesized single crystals of **1** with approximate dimension 0.01 x 0.16 x 0.18 $mm^3$ and hexagonal prism habitus were mounted on a goniometric head and investigated at 100 K with an Xcalibur3 (Oxford Diffraction) diffractometer using Mo $K_\alpha$ radiation ($\lambda$ = 0.71073 Å). The compound is isomorphic with its tetrairon(III) analogue[23] and crystallizes in trigonal space group $R\bar{3}c$ with unit cell parameters $a = b$ = 16.1435(11) Å and $c$ = 57.073(2) Å (hexagonal setting). Any possible twinning was excluded after collection of low $\theta$ frames. The Miller indices of the crystal faces were obtained by cell determination in order to orient the crystals in the EPR spectrometer (see below). The flat and more developed faces of the crystal were identified as (001) and $(00\bar{1})$ and were thus perpendicular to the trigonal symmetry axis *c*. This allowed an easy mounting for rotation around *c* and in a plane perpendicular to it.

*Electron Paramagnetic Resonance.* Single-crystal W-Band EPR measurements on **1** were performed with a Bruker E600 continuous-wave spectrometer with cylindrical cavity operating at around 94 GHz, equipped with a split-coil superconducting magnet which generates a horizontal magnetic field (Oxford Instruments). Rotation of the sample holder around a vertical axis provides the possibility for angle-resolved studies. Temperature variation was achieved with a continuous-flow cryostat (Oxford CF935), operating from room temperature down to 4.2 K. Rotation from the *c* axis to the *ab* plane was achieved by fixing the (001) face of the crystal on a lateral face of a cubic NaCl crystal, attached to the bottom of the quartz rod (see Figure S2 in supplementary material[46]). Rotation in the *ab* plane (*i.e.* around the trigonal symmetry axis) was achieved by fixing the 001 face of a single crystal on the bottom of a flat quartz rod (see Figure S3 in supplementary material[46]).

## III. RESULTS

W-band EPR spectra of **1** obtained with the static magnetic field along the trigonal *c* axis and recorded at variable temperature (6-40 K) are presented in Figure 1 in their standard derivative form. At the lowest investigated temperature seven main lines are observed, which can be attributed to $\Delta M_S = 1$ transitions between the lowest lying $M_S$ levels of the $S = 6$ ground multiplet.

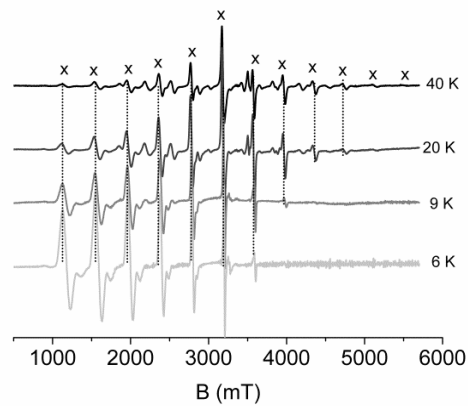

FIG. 1 Temperature dependence of the W-band EPR spectrum of **1** obtained with static field applied along the **c** crystallographic axis. The crosses and the dotted lines evidence the 12 signals of the ground S = 6 state.

Interestingly, even at this temperature, additional weaker signals are visible in between the main lines. On increasing temperature their relative intensity increases, suggesting that they originate from transitions within the first-excited spin multiplets with $S = 5$; at 20 K, the whole sets of 12 and 10 lines expected for the $S = 6$ and $S = 5$ states are observed. A further set of evenly-spaced lines, barely visible at 20 K, become more pronounced at the highest measured temperature (40 K), so that we attribute them to the second set of excited multiplets with $S = 4$. A first estimation of the axial component of the zfs for the different multiplets was made by plotting the resonance fields of the $|M_s\rangle \to |M_{s+1}\rangle$ transitions as a function of $M_S$ quantum number (Figure 2) and using a GSA based on the axial Hamiltonian:

$$\hat{H}_{ax} = \mu_B g_z B_z \hat{S}_z + D\hat{S}_z^2 + B_4^0 \hat{O}_4^0 \qquad (1)$$

where $\hat{O}_4^0$ is the fourth-order axial Stevens operator.[47] Solving Equation (1) yields the following expression for the resonance fields:

$$B_{res}(M_s) = \frac{\hbar\omega}{\mu_B g_z} - \frac{(2D - 2330 B_4^0)M_s + 140 B_4^0 M_s^3 + 210 B_4^0 M_s^2 + (D - 1200 B_4^0)}{\mu_B g_z} \qquad (2)$$

which provided the best fit parameters: $S = 6$, $g_z = 2.007 \pm 0.002$, $D = -0.1845 \pm 0.0007$ cm$^{-1}$, $|B_4^0| < 5 \times 10^{-7}$ cm$^{-1}$; $S = 5$, $g_z = 2.002 \pm 0.003$, $D = -0.155 \pm 0.001$ cm$^{-1}$; $|B_4^0| < 5 \times 10^{-7}$ cm$^{-1}$. The negative $D$ parameters indicate that **c** is an easy magnetic axis, as expected for a SMM, and that **ab** is an hard magnetic plane. The small value of $B_4^0$ for both multiplets is in agreement with the almost perfect linearity of the two plots. In this approach no reliable estimates could be obtained for $S = 4$ due to the small number of observed transitions.

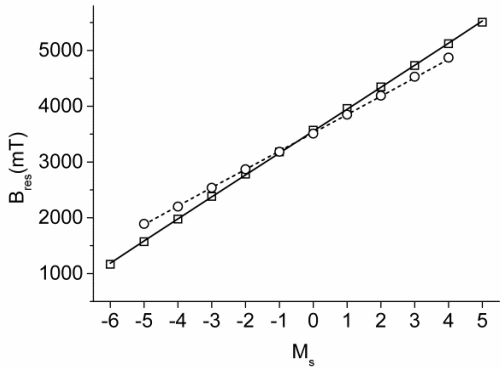

FIG. 2. Experimental resonance fields along the **c** crystallographic axis for S = 6 (squares) and S = 5 (circles), and best fits obtained using eq. 2.

Rotation of the crystal away from the **c** axis expectedly results in a reduction of the field range spanned by the spectrum, which reaches its minimum extension close to the magic angle (Figure 3). The observed behavior is in agreement with the results reported in Fig. 2, which indicate only very weak contributions from fourth (and higher order) axial terms of the Spin Hamiltonian. In this case, the angular dependence of the resonance fields is expected to approximately follow the perturbative expression:

$$B_{res}(M_S) = \frac{\hbar\omega}{\mu_B g} + \frac{(3\cos^2\theta - 1)(2M_S + 1)D}{2\mu_B g} \qquad (3)$$

which holds exactly for exclusive second-order axial anisotropy and in the strong field limit. It is immediately evident from Fig. 3 that on moving from $\theta = 0°$ to $\theta = 90°$ a relevant broadening of the EPR lines occurs, so that the spectrum in the **ab** plane is much less resolved than in the axial direction. As a consequence, the lines observed when the field is applied at large angles from the easy axis cannot be assigned by simple inspection and a complete analysis based on eq. (3) must be abandoned. It is also immediately evident that even for small values of $\theta$, (3) is not holding, indicating that the strong field limit is not fulfilled and a complete simulation has to be considered (See Figure S4 in supplementary material[46]).

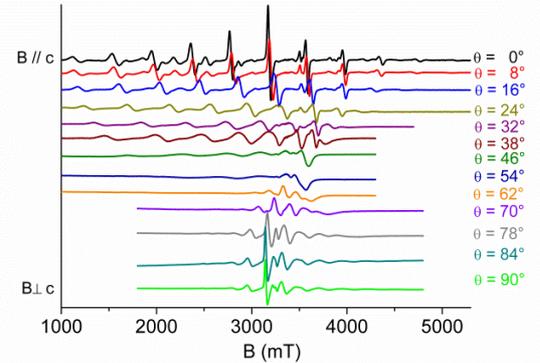

FIG. 3 (color online) Angular dependence of the W-band EPR spectrum of **1** at 20 K when rotating the crystal from the **c** crystallographic axis ($\theta = 0°$) to the **ab** plane ($\theta = 90°$, $\varphi = 15°$; here, $\varphi$ is the angle between the applied static field and the Fe-Cr direction)

In Figure 4 we present the angular dependence of the spectra recorded at 20 K by applying the static field in the *ab* plane at different angles (φ) from the Fe-Cr direction. Due to the extremely weak angular dependence of the resonance fields the intensity of the spectra was plotted in a bidimensional graph with a color intensity scale, which allows evidencing a 60° periodicity of some specific resonances (see Figure S5 in supplementary material[46] for details of a couple of field regions). This confirms the expected threefold symmetry and indicates that the rotation was correctly performed around *c* with a negligible misalignment (< 1°).

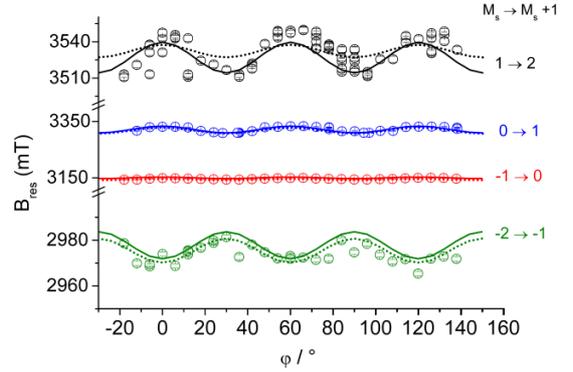

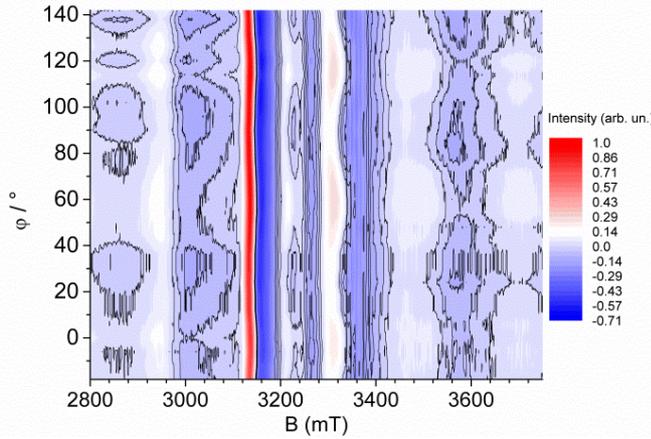

FIG. 4 (color online) Bidimensional plot of the angular dependence of W-band EPR spectra measured in the **ab** plane (i.e. perpendicular to the trigonal axis). At φ = 0° the static field is parallel to the Fe-Cr direction.

FIG. 5 (color online) Experimental (circles) and calculated angular dependence of the central $|M_S\rangle \to |M_S+1\rangle$ transitions in the **ab** plane. Continuous and dotted lines provide the resonance fields obtained using the GSA and the MSA, respectively. See text for corresponding best fit parameters.

For a $D_3$ symmetric molecule the complete giant spin Hamiltonian up to the sixth order is:

$$\hat{H}_{D_3} = \mu_B \mathbf{B} \cdot \mathbf{g} \cdot \hat{\mathbf{S}} + D\hat{S}_z^2 + B_4^0 \hat{O}_4^0 + B_4^3 \hat{O}_4^3 + B_6^3 \hat{O}_6^3 + B_6^6 \hat{O}_6^6$$

(4)

In the following discussion we will focus on the angular dependence of the transitions occurring around 2980, 3146, 3330 and 3530 mT. On the basis of the above estimates of axial zfs parameters, these bands can be unequivocally attributed to resonances within the ground $S = 6$ multiplet (the assignment to different $|M_s\rangle \to |M_s+1\rangle$ transitions is provided in Figure 5). A first relevant point to be noted is that the lowest-field transition displays an angular modulation with opposite phase as compared to the remaining three transitions. This is a clear indication that the observed periodicity cannot be ascribed to a local lowering of the $D_3$ *molecular* symmetry while preserving trigonal crystal symmetry, *i.e.* to a 3-fold symmetric distribution of rhombic anisotropies, as previously suggested for the corresponding Fe$_4$ derivative.[41] In our case, the observed angular dependence was then firmly attributed to the presence of trigonal anisotropy terms in the zfs interactions.

As a first step toward an accurate determination of $B_k^q$ parameters with $q = 3$ and 6 we performed sample calculations to test the effect of each term on the resonance field for the examined transitions. It turned out that the experimentally observed relative phases are correctly reproduced even by introducing a $B_6^6 \hat{O}_6^6$ term only. However, retrieval of the correct modulation amplitudes and resonance fields requires the introduction of both $B_4^3 \hat{O}_4^3$ and $B_6^3 \hat{O}_6^3$ terms. We note that while the sign of $B_6^6$ is directly related to the phase of angular modulation, the absolute signs of $B_k^3$ have no effect and only their relative signs could be determined from the available data. Angle-dependent measurements away from the **ab** plane would resolve this ambiguity, but are unfeasible owing to the crystal morphology. This behavior is directly related to the form of Stevens operators with odd $q$, which contain contributions from odd powers of $S_z$.[47] For the same reason these terms in principle affect the position of the parallel transitions too. Therefore, the angle-dependent resonance fields in the **ab** plane and those along the easy axis for the $S = 6$ state were simultaneously fitted using full diagonalization of the Spin Hamiltonian matrix. The best-fit simulations presented in Figure 5 were obtained with the $S = 6$ parameters gathered in Table 1.

The complete simulation of the easy-axis spectrum (Figure 6) with the correct relative intensity of the lines at different temperatures required the inclusion of a doubly-degenerate excited $S = 5$ state lying 33 K above in energy and of a triply-degenerate $S = 4$ state at 66 K, weighted according to their degeneracy and thermal population (evaluated using Boltzmann distribution). The energies of the two excited multiplets, and the degeneracy imposed by the trigonal symmetry of **1**, are in excellent agreement with magnetic susceptibility data reported for related $Fe_3Cr$ derivatives.[42,44] The broadening of the lines at the extremes of the spectra, which is often observed in the EPR spectra of SMMs,[11,31,48,49] was attributed to a distribution in the axial zfs parameters $D$, and could be correctly reproduced assuming for each multiplet a specific distribution width $\sigma_D$ (the complete set of best fit parameters can be found in Table 1).

TABLE I Best-fit parameters obtained from the simulation of the EPR spectra within the GSA for the three lowest multiplets of **1**.

|  | $S = 6$ | $S = 5$ | $S = 4$ |
|---|---|---|---|
| $g_z$ | $2.008 \pm 0.001$ | $2.008 \pm 0.001$ | $2.012 \pm 0.002$ |
| $g_{x,y}$ | $2.0131 \pm 0.001$ | - | - |
| $D$ / cm$^{-1}$ | $-0.1845 \pm 0.0005$ | $-0.1554 \pm 0.001$ | $-0.105 \pm 0.002$ |
| $B_4^0$ / cm$^{-1}$ | $+(2.0 \pm 0.1) \times 10^{-7}$ | | |
| $B_4^3$ / cm$^{-1}$ | $\pm(3.0 \pm 0.5) \times 10^{-4}$ | - | - |
| $B_6^3$ / cm$^{-1}$ | $\mp(1.0 \pm 0.1) \times 10^{-5}$ | - | - |
| $B_6^6$ / cm$^{-1}$ | $+(5.5 \pm 0.5) \times 10^{-7}$ | - | - |
| $\Delta B_{pp}$ [a] | 30 mT | 55 mT | 55 mT |
| $\sigma_D$ [a] | 100 MHz | 300 MHz | - |
| $\Delta E$ [a] | - | 33 K | 66 K |

[a] $\Delta B_{pp}$ is the distance between the position of the maximum and of the minimum in the first-derivative lineshape. $\sigma_D$ is defined as the FWHM of the Gaussian distribution of the scalar parameter $D$. $\Delta E$ is the energy of the excited multiplets above the ground $S = 6$ state.

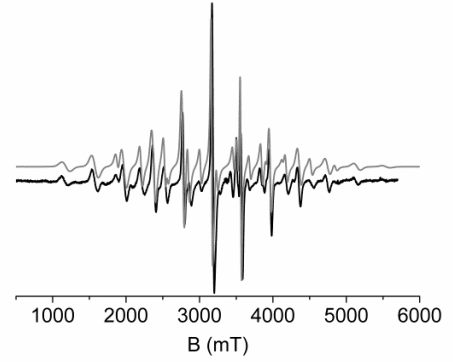

FIG. 6 Experimental (black trace) EPR spectrum measured at 40 K with static field along the **c** axis and best simulation (gray trace) obtained using parameters reported in Table1.

## IV. DISCUSSION

Single crystal W-band EPR spectra measured with static field along the easy axis **c** provided the axial zfs parameters of **1** in its ground $S = 6$ state and in its excited $S = 5$ and $S = 4$ states. To the best of our knowledge, this is the first time that the zfs parameters of two excited states in a SMM have been measured. Furthermore an accurate determination of transverse trigonal anisotropy in the ground state was possible by examining the angular dependence of EPR resonances in the **ab** plane. In particular, the simultaneous presence of both $q = 3$ and $q = 6$ terms in the giant spin Hamiltonian turned out to be necessary to correctly reproduce the observed behavior. It is essential to stress that 6$^{th}$ order anisotropy terms cannot arise from the projection of single ion anisotropies. Indeed, chromium(III) ion is an $S = 3/2$ center and its single ion anisotropy may provide neither $k = 4$ nor $k = 6$ contributions (since $k \leq 2S$); in much the same way high spin iron(III) has no 6$^{th}$ order single ion terms, being an $S = 5/2$ ion. It has further to be noted that the use of GSA, and thus of Eq. (4), to describe both ground and excited states apparently implies that the corresponding zfs parameters are independent of each other. However, in the strong exchange approximation on which GSA relies, the second order zfs tensor of any spin state $S$ for a $Fe_3Cr$ cluster is related to the microscopic anisotropic parameters (i.e. single-ion and pairwise - dipolar and anisotropic exchange - interactions) through:

$$\mathbf{D}_S = d_{Fe}(\mathbf{D}_{Fe(1)} + \mathbf{D}_{Fe(2)} + \mathbf{D}_{Fe(3)}) + d_{Cr}\mathbf{D}_{Cr}$$
$$+ d_{Fe,Fe}\sum_{i=1}^{3(j>i)}\mathbf{D}_{Fe(i),Fe(j)} + d_{Fe,Cr}\sum_{i=1}^{3}\mathbf{D}_{Fe(i),Cr} \quad (5)$$

where $\mathbf{D}_{Fe(i)}$ and $\mathbf{D}_{Cr}$ are the single-ion anisotropy tensors, $\mathbf{D}_{Fe(i),Fe(j)}$ and $\mathbf{D}_{Fe(i),Cr}$ are the sum of dipolar and anisotropic exchange ones, while $d_{Fe}$, $d_{Cr}$, $d_{Fe,Fe}$ and $d_{Fe,Cr}$ are projection coefficients calculated according to recursive relations.[50]

To account for both the observed three-fold in-plane anisotropy and the magnitude of the axial anisotropy of the excited states, the strong exchange approximation inherent to the GSA must be abandoned and a multispin Hamiltonian (MSH) introduced:

$$\hat{H}_{MSH} = \sum_{i=1}^{3(j>i)}\hat{\mathbf{S}}_{Fe(i)}\cdot\mathbf{J}_{Fe(i),Fe(j)}\cdot\hat{\mathbf{S}}_{Fe(j)} + \sum_{i=1}^{3}\hat{\mathbf{S}}_{Fe(i)}\cdot\mathbf{J}_{Fe(i),Cr}\cdot\hat{\mathbf{S}}_{Cr}$$
$$+ \mu_B\sum_{i=1}^{3}\mathbf{B}\cdot\mathbf{g}_{Fe(i)}\cdot\hat{\mathbf{S}}_{Fe(i)} + \mu_B\mathbf{B}\cdot\mathbf{g}_{Cr}\cdot\hat{\mathbf{S}}_{Cr} \quad (6)$$
$$+ \sum_{i=1}^{3}\hat{\mathbf{S}}_{Fe(i)}\cdot\mathbf{D}_{Fe(i)}\cdot\hat{\mathbf{S}}_{Fe(i)} + \hat{\mathbf{S}}_{Cr}\cdot\mathbf{D}_{Cr}\cdot\hat{\mathbf{S}}_{Cr}$$

In Equation (6), $\hat{\mathbf{S}}_{Fe(i)}$ ($i$ = 1,2,3) and $\hat{\mathbf{S}}_{Cr}$ are the spin operators for the iron and chromium centers, while $\mathbf{J}_{Fe(i),Fe(j)}$ and $\mathbf{J}_{Fe(i),Cr}$ represent the interaction tensors within iron-iron and iron-chromium pairs, respectively, containing both isotropic exchange and dipolar contributions. As before $\mathbf{D}_{Fe(i)}$ and $\mathbf{D}_{Cr}$ are the zfs tensors of the iron and chromium sites, whose g-matrices are indicated by $\mathbf{g}_{Fe(i)}$ and $\mathbf{g}_{Cr}$, respectively. Following the usual conventions for $D_3$ symmetry, we chose the molecular reference frame (XYZ) with Z along the threefold symmetry axis, Y along the Cr-Fe(1) direction and X orthogonal to Y and Z. The orientation of each local tensor eigenframe (xyz) in the molecular frame was then specified in terms of its Euler angles $\alpha$, $\beta$, $\gamma$, (ZYZ convention).[47,51] The $D_3$ molecular symmetry imposes a number of constraints on the tensors/matrices appearing in Equation (6). For instance, $\mathbf{D}_{Cr}$ and $\mathbf{g}_{Cr}$ must be axial along Z and a principal direction of $\mathbf{D}_{Fe(i)}$ and $\mathbf{g}_{Fe(i)}$ must lie along the twofold axis joining Fe(i) with Cr. In addition, the three $\mathbf{D}_{Fe(i)}$ tensors and the three $\mathbf{g}_{Fe(i)}$ matrices must be related by a threefold rotation around Z, with similar relationships holding for $\mathbf{J}_{Fe(i),Fe(j)}$ and $\mathbf{J}_{Fe(i),Cr}$ tensors (see Figure 7).

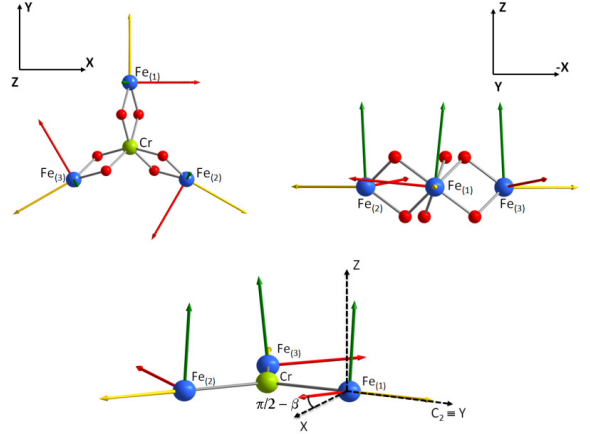

FIG. 7 (Color online) Arrangement of single-ion anisotropy tensors for the iron(III) ions in Fe$_3$Cr with respect to the molecular reference frame (XYZ). The single-ion tensors are related by a threefold rotation along Z and have a principal axis along the Cr-Fe direction, which is a $C_2$ symmetry axis. Red arrow: hard axis; yellow arrow: intermediate axis; green arrow: easy axis.

For simplicity, the isotropic exchange coupling constants $J_{Fe-Fe}$ and $J_{Fe-Cr}$ were held fixed to the values reported in Ref. [44] while dipolar contributions were calculated within the point-dipole approximation. Due to the aforementioned symmetry-imposed constraints, the number of free parameters to be determined in Eq. (6) is actually seven, hence lower than in the GSA (see Table 1).

Owing to the dimension of the Hilbert space (864x864) and to the presence of low-lying excited states, the direct simulation of EPR spectra using Equation (6) was found quite demanding. Indeed, significantly populated levels belonging to different multiplets undergo extensive crossings at relatively low field (see Figure S6 in supplementary material[46]). To reduce the complexity of the problem, the analysis was then restricted to the levels involved in the experimentally observed resonances. These levels were identified by a preliminary analysis of the field dependent energy pattern using anisotropy parameters from previous works. The axial ($D_{Cr}$, $D_{Fe}$) and rhombic ($E_{Fe}$) single-ion anisotropy parameters were set to the values found in an isostructural Ga$_4$ derivative doped with chromium(III) and iron(III) ions:[42] $D_{Cr}$ = 0.46 cm$^{-1}$, $D_{Fe}$ = 0.656 cm$^{-1}$ and $E_{Fe}$ = 0.089 cm$^{-1}$, with $g_{Fe}$ = 2.003 (as expected for a $^6$S ion) and $g_{Cr}$ = 1.98, as commonly observed for chromium(III).[47] Noticeably, since all constituent ions have a hard-axis type anisotropy ($D > 0$), the observed easy-axis molecular anisotropy requires highly non-collinear $\mathbf{D}_{Fe(i)}$ tensors. Because of the aforementioned restrictions imposed by symmetry, the line joining Cr with Fe(i) can correspond either to the *hard* (z), to the *easy* (y) or to the *intermediate* (x) axis of the $\mathbf{D}_{Fe(i)}$ tensor. The first possibility was ruled out as it results in a 30° phase mismatch with respect to the observed resonance field modulation in the ***ab*** plane (*i.e.* $B_6^6 < 0$ in the GSA). This conclusion, which fully supports previous studies,[36,40,42] implies that the hard axis (z) of each $\mathbf{D}_{Fe(i)}$ tensor is normal to the corresponding Cr-Fe(i) direction, with no symmetry-imposed restriction on the angle β between z and Z.

TABLE 2 Best-fit parameters obtained from the simulation of the EPR spectra within the MSA. $J_{Fe-Fe}$ and $J_{Fe-Cr}$ were kept fixed to the values determined by magnetic susceptibility studies. The single ion tensor for Fe(1) expressed in the molecular reference frame is reported in the second column, the corresponding tensors for Fe(2) and Fe(3) being obtained by application of the appropriate rotation of ± 120° around the Z axis.

|  |  | *Fe(1) single ion tensor in XYZ (cm$^{-1}$)* |  |
|---|---|---|---|
| $g_{Fe}$(isotropic) | 2.005 ± 0.001 | $D_{XX}$ | 0.486 |
| $D_{Fe}$ / cm$^{-1}$ | 0.738 ± 0.003 | $D_{XY}$ | 0 |
| $E_{Fe}$/cm$^{-1}$ | 0.064 ± 0.002 | $D_{YY}$ | -0.182 |
| α | 0° | $D_{XZ}$ | 0 |
| β | 85° (95°) ± 1° | $D_{YZ}$ | 0.0696 |
| γ | 90° | $D_{ZZ}$ | -0.304 |
| $g_{Cr,\perp}$ | 1.968 ± 0.001 |  |  |
| $g_{Cr,//}$ | 1.978 ± 0.001 |  |  |
| $D_{Cr}$/cm$^{-1}$ | 0.470 ± 0.005 |  |  |
| $J_{Fe-Fe}$/cm$^{-1}$ | -0.34 |  |  |
| $J_{Fe-Cr}$/cm$^{-1}$ | 13.65 |  |  |

However, γ can have only two possible values, 0 or 90°, depending on whether y or x is found along Cr-Fe(i). Additional guidance in better defining the orientation of $\mathbf{D}_{Fe(i)}$ tensors is provided by projection formulae. According to Eq. (7) in Ref. [42], the observed D parameter in the S = 6 state requires the ZZ-component of $\mathbf{D}_{Fe(i)}$ to take the value $D_{ZZ} \cong -0.30$ cm$^{-1}$. This is very close to the largest negative component that can be reached with the adopted $D_{Fe}$ and $E_{Fe}$ values ($D_{ZZ}$ = -0.31 cm$^{-1}$), suggesting that the local easy axis y is roughly parallel to Z, *i.e.* $\beta \cong 90°$ and $\gamma = 90°$. It is however apparent that the same molecular D can also be retrieved by setting $\beta \cong 90°$ and $\gamma = 0°$, provided that $D_{Fe}$ and $E_{Fe}$ are adjusted so as to afford the required $D_{ZZ}$. We could resolve this ambiguity by examining the angular variation of resonance fields in the hard plane predicted by the two arrangements. Indeed, if $D_{ZZ}$ is kept constant to allow for a correct reproduction of parallel spectra, for γ = 90° the modulation amplitudes become larger as rhombicity is reduced, while the reverse holds for γ = 0°. This clearly indicates that the angular dependence of resonance fields in the ***ab*** plane, and thus the magnitude of the transverse trigonal anisotropy, is directly related to differences in the components of $\mathbf{D}_{Fe(i)}$ along X and Y. The arrangement with the *easy* axis y along Cr-Fe(i) invariably results in modulation amplitudes larger than observed, thereby ruling out the γ = 0° option. On these grounds a very good reproduction of the hard-plane resonance fields, both compared to the experimental data and to GSA (Figure 4, dotted lines), was obtained by using the set of parameters gathered in Table 2.

The same set correctly reproduces the transitions observed in parallel spectra, both for the formally $S = 6$ ground state and for the two $S = 5$ excited states (see Fig. S7 in supplementary material[46]). Indeed, application of Eq. (5) using the single ion tensors reported in Table 2 and the appropriate projection coefficients results in an estimate of $D_6 = -0.189$ cm$^{-1}$, $D_5 = -0.159$ cm$^{-1}$ and $D_4 = -0.117$ cm$^{-1}$, in good agreement with the results obtained in the GSA. It is worth noting that in the framework of the MSA different resonance fields are calculated for the two $S = 5$ states. Indeed, **i**nclusion of single ion anisotropy terms lifts the degeneracy imposed to the two formally $S = 5$ excited states by the threefold symmetry of the exchange coupling pattern. The experimental resolution along this direction is however not enough to discriminate between signals deriving from the two states. On the other hand, the aforementioned degeneracy lifting explains some subtle features observed in the perpendicular spectrum. Two lines with a temperature dependence characteristic of transitions within excited states are detected around 3220 mT and 3250 mT (see Fig. S8 in supplementary material[46]). They show distinctly different angular dependences, the second one being essentially angle independent. This behavior cannot be reproduced within the GSA, unless very different zfs parameters are assigned to the two $S = 5$ states. On the contrary the MSA correctly predicts the values of resonance fields, their temperature dependence and different angular dependence (see Figure S9 in supplementary material[46]).

In much the same way as in the GSA, some ambiguity on the parameter values remains unresolved. For a correct simulation of the spectra it is necessary that $\beta \neq 90°$, which lowers the symmetry of the Hamiltonian from $D_{6h}$ (for $\beta = 90°$) to $D_{3d}$ (for $\beta \neq 90°$) and allows for nonzero values of the $B_k^3$ parameters ($k = 4, 6$) in the GSA. However, setting $\beta$ to $\pi - \beta$ provides coincident results, due to symmetry reasons. The two options actually correspond, in the GSA, to different choices for the sign of $B_k^3$ parameters (see above). We can then conclude that the major features of trigonal anisotropy (*i.e.* the phase of the in-plane resonance field variation), as described by the $B_6^6 \hat{O}_6^6$ term in the GSA, reflect the angular modulation of single-ion anisotropy components in the **ab** plane. However, the symmetry lowering induced by the tilting of the single ion tensors out of the **ab** plane generates $B_k^3 \hat{O}_k^3$ terms ($k = 4, 6$) which are crucial to accurately explain the experimental data. As we will further show below, this is expected to have some relevant consequences on the spin dynamics of **1**.

## V. EFFECT ON TUNNEL SPLITTING

As mentioned in the Introduction, transverse anisotropy plays a key role in determining the spin dynamics of SMMs at low temperature. In particular, it has been shown in the past that TS oscillations can be observed when a transverse magnetic field is applied along the hard direction: these oscillations are a consequence of topological interferences in the tunneling pathways, also known as Berry Phase Interference, and have been employed to investigate parity effects in the QT of integer and half-integer spin systems.[10,27,52,53] The accurate spectroscopic determination of transverse anisotropy in **1** using both GSA and MSA allows exploring the consequences of trigonal symmetry on TS oscillations. The system under investigation is especially well suited for this scope. At variance with its tetrairon(III) analogue, which features an $S = 5$ ground state, **1** has an $S = 6$ ground state, and tunneling within the lowest doublet ($M_S = \pm 6$) is promoted by the transverse terms allowed in trigonal symmetry, even without application of a transverse static field. This would better evidence the field induced "diabolic points", that is, those sets of components of the applied field for which, according to the Wigner – Von Neumann theorem, exact degeneracy is observed (TS = 0) with no symmetry requirements.[54]

We then begin our analysis by focusing on the periodicity expected for the TS between the two lowest sublevels (indicated as $\Delta_{-66}$) by application of a transverse field ($B_t$) and a compensating longitudinal field needed to exactly locate the minimum of the TS. The application of a compensating field reflects the actual experimental procedure which locates the TS minimum by a sweep of the longitudinal field around zero.[10] Based on the parameters derived from EPR spectra within the GSA, two different sets of TS minima occur at two magnitudes of $B_t$ along directions $\varphi = \pm n\pi/3$ (n integer) (see Figure 8, left).

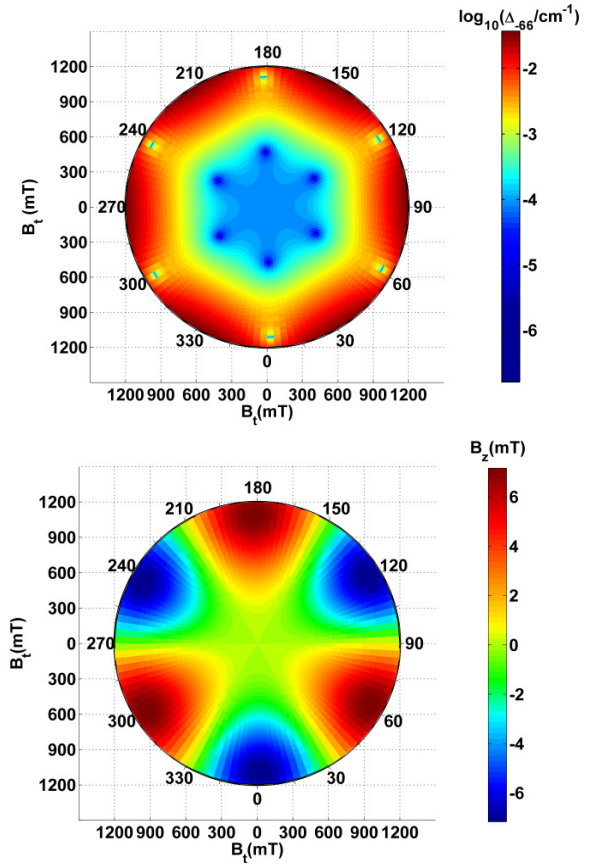

FIG. 8 (Color online) Left: Tunnel splitting periodicity between the fundamental levels $M_S = \pm 6$ on application of a transverse field of variable magnitude in the **ab** plane, calculated in the GSA. Right: Compensating longitudinal field necessary to locate the minima of the tunnel splitting reported in the left panel. Note that the sign of the compensating field depends on the choice made for the absolute sign of $B_4^3$ and, correspondingly, of $B_6^3$: the plot corresponds to $B_4^3 > 0$, $B_6^3 < 0$.

The apparent sixfold modulation of the tunneling gaps is at first sight in contrast with the trigonal symmetry of **1**. However, it has to be considered that also the compensating longitudinal field undergoes a threefold modulation in the ***ab*** plane (Figure 8, right). From an experimental point of view this should result in the observation of an hexagonal symmetry of the TS variation, and thus of QT efficiency, in the hard plane. On the contrary, a trigonal symmetry is expected when the investigation is performed out of plane, since the longitudinal component of the field will differently affect, in this case, minima occurring every $\pi/3$ (see Figure S10 in supplementary material[46]). This behavior can be considered as the signature of the presence of $B_k^3 \hat{O}_k^3$ terms in the GSH, since they contain odd contribution of $\hat{S}_z$ which act as an effective internal longitudinal field.[20] Indeed, simulations obtained neglecting $B_k^3 \hat{O}_k^3$ terms do not show appreciable modulation of the compensating field, in agreement with the overall higher symmetry of the Spin Hamiltonian.

It is interesting to compare the behavior predicted on the basis of GSA with the one expected within the MSA (Figure 9). In the latter case TS minima are also observed along directions $\varphi = \pm n\pi/3$, but at fields appreciably different ($B_t$ = 305 and 915 mT) from those predicted by the GSA ($B_t$ = 474 and 1114 mT). Furthermore, for both minima in MSA the value of the longitudinal compensating field ($|B_z|$=6 x $10^{-2}$ mT and $|B_z|$=1.6 mT respectively) is smaller than that obtained in the GSA ($|B_z|$=1.35 mT and $|B_z|$=7.15 mT). Finally, we note that for small transverse field the GSA predicts somewhat larger TS values than the MSA. This is in line with previous results obtained by some of us in the simplified multispin modeling of the tetragonal $Mn_{12}tBuAc$ system,[19] but contrasts with other findings on different systems.[17]

As a whole these results evidence that the two different approaches, even when providing extremely high quality reproduction of EPR spectra, may result in somewhat different predictions of the field-dependent spin dynamics. This may be attributed to the fact that GSA high order parameters provides only a phenomenological description of the transverse anisotropy, without any assumption in term of their physical origin. For this reason, while the GSA model can accurately describe spectroscopic properties of exchange coupled systems, it may provide inaccurate prediction as for the TS behavior, which is extremely sensitive to differences in the energy eigenstates and to the mixing between different multiplets, neglected in this approach.

On the other hand, MSA is in principle more rigorous than GSA and provides a more satisfactory description of magnetic anisotropy by considering the details of single-ion anisotropies and spin-spin interactions. This allows to explain more subtle properties of the system and to trace back the origin of high order anisotropy in GSA to the non-collinearity of single ion tensors in MSA. However, it often relies on a large number of parameters whose univocal determination may be difficult in the absence of further experimental information. Noticeably, in the system studied here the high symmetry of the cluster and of the ion sites reduces the number of free parameters to below that required by the GSA. However even in the case of **1**, for which an accurate determination of the single ion tensors could be obtained, some potentially relevant contribution to the anisotropy, such as anisotropic exchange or Dzyaloshinskii–Moriya interactions,[50,55] were neglected, thus leaving some degree of uncertainty about the predicted TS modulation. In other words, the MSA parameterization we have used is the simplest model able to account for the spectroscopic set of data.

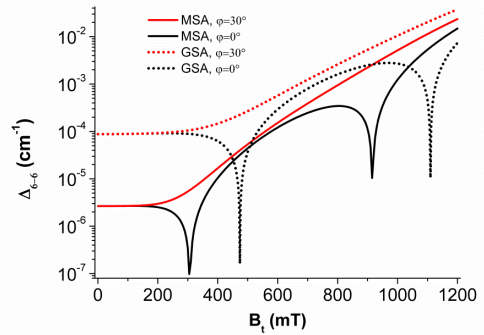

FIG. 9 (Color online) Transverse field dependence of tunnel splitting within the ground doublet of **1** calculated using the spin Hamiltonian parameters that best reproduce EPR spectra in the GSA (dotted lines) and in the MSA (continuous lines). The transverse field is applied at $\varphi = 0°$ (black traces) and $\varphi = 30°$ (red traces) in the **ab** plane.

The comparison between the GSA and MSA approaches reveals most striking differences in the transverse field dependence of the tunnel splitting of the low lying doublets. An investigation of the low temperature spin dynamics could actually clarify the influence of the neglected anisotropic contributions (anisotropic exchange and Dzyaloshinskii–Moriya ones) on the Single Molecule Magnet behavior, which is still an open issue.[51,56–60] Unfortunately, preliminary low temperature spin dynamics measurements[61] showed that the tunnel rate in **1** is too fast for a reliable estimate of the TS based on standard Landau-Zener method.[10,59]

## VI. CONCLUSIONS

We presented a detailed single crystal W-band EPR investigation of **1**, a tetranuclear SMM with crystallographically imposed $D_3$ symmetry and a ground $S = 6$ state. Accurate axial parameters were obtained for both the ground state and the lowest excited states, $S = 5$ and $S = 4$. The angular dependence of the spectra in the hard plane allowed us to firmly establish the presence of high order transverse anisotropy terms that determine a 60° periodicity of the resonance fields. Thanks to the sensitivity of single crystal W-band EPR, the corresponding giant spin Hamiltonian parameters were determined for the first time in a SMM. The spectral behavior was further reproduced using a complete MSA, starting from previously reported results on Fe and Cr-doped $Ga_4$ analogues. By comparing the results obtained in the two approaches, we found that trigonal anisotropy originates from the breaking down of the strong exchange approximation. In particular, it directly reflects the structural features of the cluster, *i.e.* the relative orientation of the single ion anisotropies and the different single ion anisotropy components in the hard plane of the cluster.

Finally, since the transverse anisotropy terms play a key role in the quantum tunneling regime, we investigated their effect on the tunnel splitting within the ground doublet. Although accounting equally well for the available EPR data, the two descriptions (GSA and MSA) yielded somehow different predictions. Despite the failure of preliminary low temperature spin dynamics measurements to clarify this point, further attempts to measure Berry Phase Interference patterns will be performed in the future. Indeed, **1** offers some advantages as compared with the $Mn_3$ complex investigated by Hill and co-workers:[62] molecules within the crystal are iso-oriented, and dilution of **1** in a diamagnetic isomorphous $Ga_4$ matrix can be envisaged to reduce intermolecular dipolar interactions, as recently reported for this family of molecules.[39]

## ACKNOWLEDGEMENTS

Financial support from the European Research Council through the Advanced Grant MolNanoMas (grant no. 267746), Brazilian CNPq (Conselho Nacional de Desenvolvimento Científico e Tecnológico, grant no. 201817/2009-8) is gratefully acknowledged. P.T. and J.F.S. thank Ministero dell'Istruzione, dell'Università e della Ricerca Italiana, CNPq and CAPES (Coordenação de Aperfeiçoamento de Pessoal de Nível Superior, Brazil) for fellowships.